\documentclass[12pt,epsf]{article}
\usepackage{amssymb,amsmath}
\usepackage{graphicx}

\newcommand{\be}{\begin{equation}}
\newcommand{\ee}{\end{equation}}
\newcommand{\bea}{\begin{eqnarray}}
\newcommand{\eea}{\end{eqnarray}}
\newcommand{\beas}{\begin{eqnarray*}}
\newcommand{\eeas}{\end{eqnarray*}}
\newcommand{\ba}{\begin{array}}
\newcommand{\ea}{\end{array}}

\newcommand{\n}{\nu}

\newcommand{\tr}{\mathrm{Tr}}

\newcommand{\diag}{\mathrm{diag}}
\newcommand{\Ncal}{\mathcal{N}}

\newcommand{\nbox}{{\,\lower0.9pt\vbox{\hrule \hbox{\vrule height 0.2 cm \hskip 0.19 cm \vrule height 0.2 cm}\hrule}\,}}

\def\href#1#2{#2}

\begin{document}
\begin{titlepage}
\hfill
\vbox{
    \halign{#\hfil         \cr
           arXiv:0705.4308 \cr
           } 
      }  
\vspace*{20mm}
\begin{center}
{\Large \bf Coarse-Graining the \\ Lin-Maldacena Geometries}

\vspace*{15mm}
\vspace*{1mm}

Hsien-Hang Shieh, Greg van Anders, and Mark Van Raamsdonk

\vspace*{1cm}

{Department of Physics and Astronomy,
University of British Columbia\\
6224 Agricultural Road,
Vancouver, B.C., V6T 1Z1, Canada}

\vspace*{1cm}
\end{center}

\begin{abstract}
The Lin-Maldacena geometries are nonsingular gravity duals to degenerate vacuum states of a family of field theories with $SU(2|4)$ supersymmetry. In this note, we show that at large $N$, where the number of vacuum states is large, there is a natural `macroscopic' description of typical states, giving rise to a set of coarse-grained geometries. For a given coarse-grained state, we can associate an entropy related to the number of underlying microstates. We find a simple formula for this entropy in terms of the data that specify the geometry. We see that this entropy function is zero for the original microstate geometries and maximized for a certain ``typical state'' geometry, which we argue is the gravity dual to the zero-temperature limit of the thermal state of the corresponding field theory. Finally, we note that the coarse-grained geometries are singular if and only if the entropy function is non-zero.

\end{abstract}

\end{titlepage}

\vskip 1cm

\section{Introduction}

Recently, several fascinating new examples of gauge-theory / gravity duality have emerged \cite{lm} for which the field theory has a discrete highly degenerate basis of vacuum states yet we have an explicit non-singular geometry corresponding to each element of the basis.

The field theories include the Plane-Wave Matrix Model (a one-parameter deformation of the low-energy theory of D0-branes \cite{bmn}), a maximally supersymmetric 2+1 dimensional gauge theory on $S^2$ \cite{msv}, ${\cal N}=4$ SUSY Yang-Mills theory on $S^3/Z^k$, and type IIA Little String Theory on $S^5$ \cite{lm,msv,lmsvv}. Each of these theories has $SU(2|4)$ supersymmetry, which may be used to argue that the numerous classical vacuum states (reviewed in section 2) remain degenerate in the quantum theory, and in particular, must be present at strong coupling. In \cite{lm} (following \cite{llm}), Lin and Maldacena searched for supergravity solutions with the same $SU(2|4)$ symmetry, and found nonsingular solutions in one-to-one correspondence with each element of a natural basis of vacuum states for each of the field theories.\footnote{The construction is completely analogous to the construction of gravity duals to half BPS states of {\cal N}=4 SUSY Yang-Mills theory \cite{llm}. As in that case, the smooth supergravity solutions corresponding particular states can have large curvatures, and thus are only approximations to the true dual geometries which should minimize the $\alpha'$-corrected low-energy effective action.} In the following discussion and sections 2 to 5 of this paper, we focus on the example of the Plane-Wave Matrix Model (reviewed in section 2), but we discuss the other theories in detail in section 6.

While the geometries corresponding to basis vacuum states in each case are the same asymptotically, they differ even in their topology in the infrared. Since the generic vacuum state in the field theory is a linear superposition of basis elements, such a state cannot be dual to a single non-singular supergravity solution with fixed topology (assuming there are observables that can detect topology), but must simply be dual to a quantum superposition of the topologically different geometries. Similarly, generic mixed states in the field theory, such as the zero-temperature limit of the thermal state, involve microstates corresponding to many different topologies so we might expect that a gravitational dual description in terms of a single geometry is impossible. On the other hand, there are many examples of geometries believed to be dual to thermal states of field theories, and these thermal states involve enormous numbers of microstates that can be very different in the infrared. Mathur and collaborators have advocated (see \cite{mathur} for a review) that we should interpret the thermal state geometry as a coarse-grained description of the underlying microstates, just as the homogeneous configuration that we use to describe the thermal state of a gas in a box is a coarse-graining of the true microstates of the atoms. Specifically, the macroscopic description of any almost any state in the underlying ensemble of microstates is extremely close to one particular coarse-grained configuration, the thermal equilibrium state. We will see that in our case also, there is a natural way to coarse-grain (i.e. give a macroscopic description of)  geometries corresponding to typical microstates, and that most of the microstates have a coarse-grained description that is very close to a particular geometry, which we propose is the correct dual to the zero temperature limit of the thermal state. In this geometry, the complicated topological features that distinguish the individual microstate geometries are replaced by a singularity.\footnote{For a general discussion of conditions under which field theory states can be associated with semiclassical geometries, see \cite{vijaynew} in the LLM context and \cite{larjo} in the D1-D5 context.}

The details of the coarse graining procedure are described in section 3 below, but we give the essential idea here. The supergravity fields in the  Lin-Maldacena geometries are determined in terms of the potential for an axially symmetric electrostatics problem involving a certain number of parallel coaxial charged conducting disks in a background electric field. The number, locations and charges of the disks are determined by the data specifying the field theory vacuum.\footnote{The radii of the disks are determined by the other information via a constraint.} We will find that typical field theory vacua correspond to electrostatics configurations with a large number of closely spaced disks whose radii are very small compared with the separation between the disks. At large $N$, such a configuration has a natural coarse-grained description as a smoothly varying charge distribution on the axis.  Inserting the potential arising from this coarse-grained configuration into the Lin-Maldacena supergravity solution, one finds a singular geometry. Since all of the nontrivial topological features are associated with the regions between the disks in the electrostatics configurations (these regions map to topologically non-trivial throats in the supergravity solutions), we see that the complicated topologies that characterize individual microstates are replaced by a singularity in the coarse-grained description.\footnote{We should note however, that for the case of closely spaced disks, the supergravity approximation is not valid for the region between the disks, so the classical topological features that we are discussing should be understood to be replaced by some stringy analogue.}

A completely analogous coarse-graining has been discussed \cite{babel,buchel,shepard,silva} for the half-BPS sector of ${\cal N}=4$ SUSY Yang-Mills theory. There, the microstate geometries are the type IIB LLM geometries \cite{llm}, constructed in terms of droplets of a two-dimensional incompressible fluid, and the coarse-grained description allows for configurations with arbitrary density of the fluid between zero and the maximal density. One significant difference is that all of the states we consider are ground states for the field theory, whereas the LLM discussion relates to a special class of excited states with energy equal to an R-charge.

As emphasized in \cite{babel}, a given coarse-grained configuration provides an approximation to a very large number of microstates, just as in the thermodynamic description of ordinary physical systems. Further, there is one preferred coarse-grained configuration, analogous to the thermal equilibrium state, which is very close to the coarse-grained description of almost any randomly chosen microstate. For the type IIB LLM geometries, the geometry corresponding to this preferred state was determined in \cite{babel} and dubbed the ``hyperstar'' geometry. In section 3 of this paper, we determine the corresponding geometry for the Plane-Wave Matrix Model. In our case, the ensemble of microstates we consider is just the set of vacuum states, or alternately the set of states that contribute (each with equal weight) to the $T \to 0$ thermal state density matrix. Thus, we propose that our preferred geometry is the $T \to 0$ limit of the  geometry dual to the thermal state of the field theory. In section 5, we also derive geometries corresponding to the preferred states in other restricted ensembles, analogous to the type IIB superstar \cite{mt}, and discuss thermal geometries for the remaining $SU(2|4)$ symmetric theories in section 6.

As for an ordinary thermodynamic system, the thermal states we derive should maximize an entropy functional that measures the number of microstates nearby an arbitrary coarse-grained configuration. In section 4, we derive such an entropy functional, and find that it may be written simply in terms of the data that specify the geometry. We find that this functional is indeed maximized by the thermal state geometry of section 4. Further, we note that for all the coarse-grained configurations, those for which the entropy functional vanishes are the ones that coincide the original non-singular microstate geometries. On the other hand, configurations with non-zero entropy are necessarily singular.

In the general proposal by Mathur and collaborators, black hole geometries with horizons are to be understood as coarse-grained descriptions of underlying horizon-free microstate geometries. In the present setup, the coarse graining leads to geometries with naked singularities uncloaked by horizons, but this is to be expected since the number of microstates in our case is not large enough to give a classical finite-area horizon in the supergravity limit. It may be that a horizon develops as we move from the supergravity approximation to solutions minimizing the full low-energy effective action, but, as we will see, realizing this would necessarily involve understanding both $\alpha'$ and string loop corrections.

\section{The $SU(2|4)$ symmetric matrix quantum mechanics and the dual Lin-Maldacena Geometries}

In this section, we review the Plane-Wave matrix model, its vacua, and the dual geometries constructed by Lin and Maldacena. The other $SU(2|4)$ symmetric field theories are discussed in section 6. We will see that each of these theories has a large degeneracy of vacuum states at the classical level. This degeneracy remains at the quantum level, since the representation theory of $SU(2|4)$ does not allow for states with arbitrarily small non-zero energies, and therefore does not allow the zero-energy states in the classical limit of the theory to receive corrections to their energy \cite{dsv2,kp}.

\subsection{The Plane-Wave Matrix Model}

The Plane-Wave Matrix Model \cite{bmn} is a massive deformation of the supersymmetric matrix quantum mechanics describing decoupled low-energy D0-branes in flat space.\footnote{This is similar to the Polchinski-Strassler deformation of ${\cal N}=4$ SUSY Yang-Mills theory \cite{ps}, but in this case, we preserve all 32 supersymmetries.} It is described by a dimensionless Hamiltonian
\bea
\label{PPmatrix2}
H &=& \tr \left( {1 \over 2} P_A^2 + {1 \over 2} (X_i/3)^2 + {1 \over 2} (X_a/6)^2
+ {i \over 8} \Psi^\top \gamma^{123} \Psi \right. \cr
&& \qquad \left.+  {i \over 3} g \epsilon^{ijk} X_i X_j X_k - {g \over 2} \Psi^\top \gamma^A [X_A, \Psi] - {g^2 \over 4} [X_A, X_B]^2 \right) ,
\eea
where $A=1, \dots , 9$, $i=1, \dots ,3$, and $a=4, \dots ,9$. Here, the scalars $X_A$ and 16-component fermions $\Psi$ are hermitian $N \times N$ matrices, and $P_A$ is the matrix of canonically conjugate momenta. Apart from $N$, the size of the matrices, the theory has one dimensionless parameter $g$, such that the theory is weakly coupled for small enough $g$.\footnote{The model was introduced originally as a matrix model for M-theory on the maximally supersymmetric eleven-dimensional plane-wave. For this we are required to take a limit $N \to \infty$ with $g^2 N \sim N^4$. In the present work, we will mainly be concerned with the usual 't Hooft large $N$ limit with $\lambda$ fixed.}

For this theory, the classical vacua, each with zero energy, are described by
\[
X^a = 0 \qquad a=4,\dots,9 \qquad \qquad X^i = {1 \over 3g} J^i \qquad i = 1,2,3,
\]
where $J^i$ give any reducible representation of the $SU(2)$ algebra. These vacua are in one-to-one correspondence with partitions of $N$, since we may have in general $n_k$ copies of the $k$-dimensional irreducible representation such that $\sum_k k n_k = N$. Below, it will be convenient to represent such a partition by a Young diagram with $N$ boxes, containing $n_k$ columns of length $k$.

In the D0-brane picture, a block-diagonal configuration with $n_k$ copies of the $k$-dimensional irreducible representation is associated classically with concentric D2-brane fuzzy spheres, with $n_k$ spheres at radius proportional to $k$. On the other hand, it was argued in \cite{msv} that at sufficiently strong coupling, such a configuration is better described as a collection of concentric fivebranes, with multiplicities and radii given in terms of the numbers and lengths of columns in the dual Young diagram.\footnote{In \cite{msv}, the matrix model was discussed in the context of its conjectured description of M-theory on a plane-wave background. There, the fivebranes were M5-branes, while here we are considering a limit with fixed $\lambda$, dual to a IIA background, so the fivebranes are NS5 branes.} For general values of parameters, we can interpret the solution as a fuzzy configuration with both D2-brane and NS5-brane characteristics. This will be apparent from the dual gravitational solutions, which include throats carrying D2-brane flux and throats carrying NS5-brane flux in the infrared part of the geometry.

\subsection{Electrostatics}

The vacua of the matrix model each preserve $SU(2|4)$ symmetry. In \cite{lm}, Lin and Maldacena searched for type IIA supergravity solutions preserving the same $SU(2|4)$ symmetry (more precisely, with isometries given by the bosonic subgroup $SO(6) \times SO(3) \times U(1)$ of $SU(2|4)$). Using an ansatz with this symmetry (reproduced in appendix \ref{sgapp}), they were able to reduce the problem of finding supergravity solutions to the problem of finding axially-symmetric solutions to the three-dimensional Laplace equation, with boundary conditions involving parallel charged conducting disks and a specified background potential. Corresponding to each classical vacuum and choice of parameters, we have a specific electrostatics problem, whose solution (a potential $V(r,z)$) feeds into the equations (\ref{IIA ansatz last}) to give the dual supergravity solution. Further, the smooth supergravity solutions for which fluxes through non-contractible cycles are quantized appropriately are in one-to-one correspondence with the vacua.

For the other $SU(2|4)$ symmetric theories described in section 6, the construction differs only by a choice of boundary conditions (background potential or the presence/absence of infinite-sized conducting plates). The solution to these electrostatics problems has been discussed in \cite{inst}.

We now describe the electrostatics problem in detail and then review some general features of the dual supergravity solutions.
Common to all vacua, we have in the electrostatics problem an infinite conducting plate at $z=0$ (on which we may assume that the potential vanishes), and a background potential
\be
\label{vinf}
V_\infty = V_0(r^2 z - {2 \over 3} z^3) \; .
\ee
In addition, corresponding to a matrix model vacuum with $Q_i$ copies of the $d_i$-dimensional irreducible representation, we have conducting disks with charge $Q_i$ parallel to the infinite plate and centred at $r=0$, $z = d_i$.\footnote{Our conventions here are slightly different from the ones in \cite{lm}, as we describe in appendix A.} In order that the supergravity solution is non-singular, the radii $R_i$ of the disks must be chosen so that the charge density at the edge vanishes.

The parameters of the matrix model are related to the parameters in the electrostatics problem as $N = \sum Q_i d_i$ and $g^2 \propto 1/V_0$.

\begin{figure}
\centering
\includegraphics{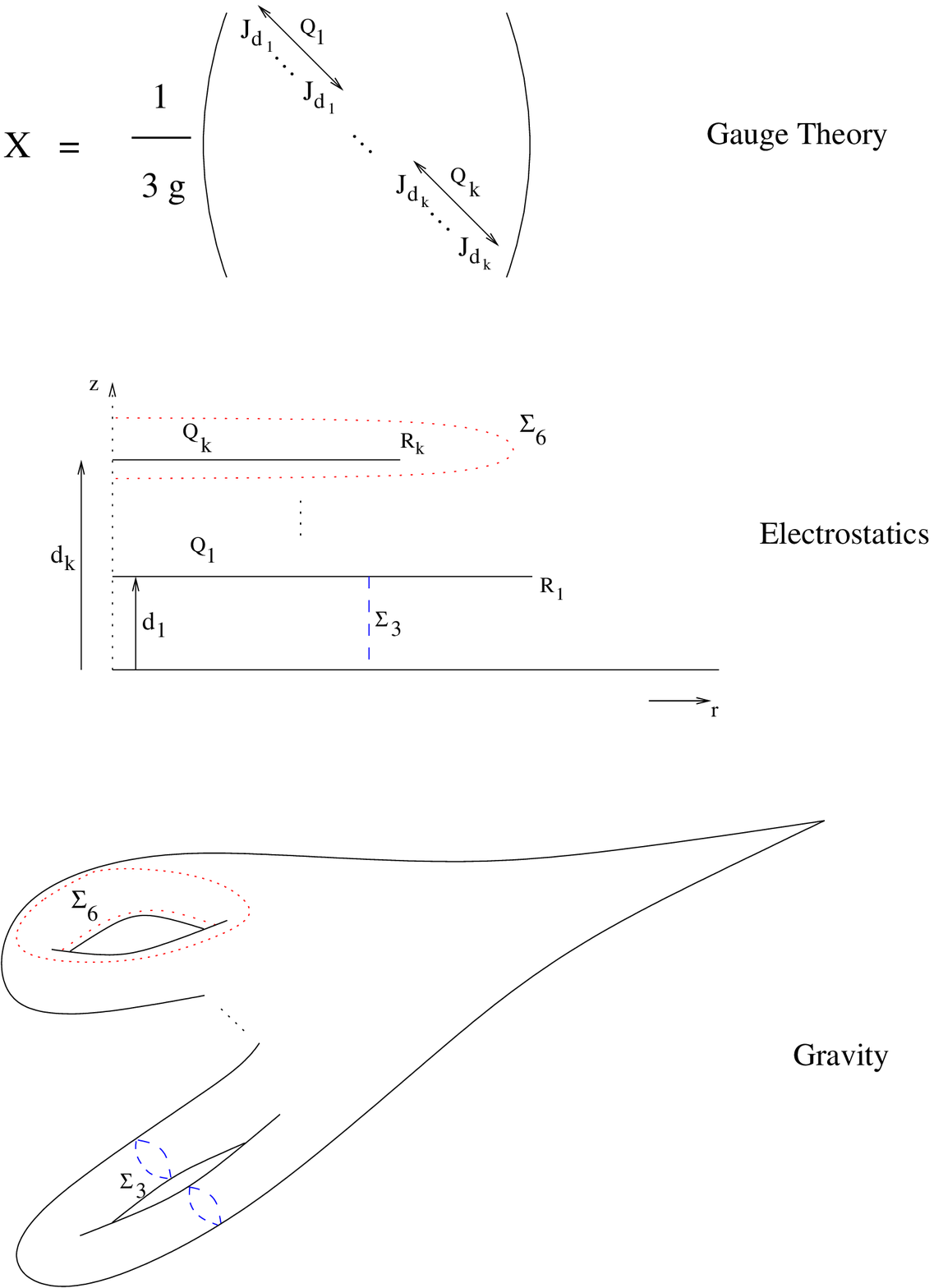}
 \caption{Mapping between matrix model vacua, electrostatics configurations, and geometries. For illustrative purposes, we have replaced the $S^2 \times S^5$s associated to each point $(r,z)$ with $S^0 \times S^0$. In the full geometry, the dotted segment maps to a submanifold $\Sigma_6$ that is topologically $S^6 \times S^2$ (simply connected) rather than the $S^1 \times S^0$ shown here. Similarly, the dashed segment maps to a submanifold $\Sigma_3$ that is topologically $S^5 \times S^3$ rather than the $S^0 \times S^1$ here.}
\label{fig1}
\end{figure}

\subsection{Gravity Duals}

The coordinates $r$ and $z$ in the electrostatics problem form two of the nine spatial coordinates in the geometry. In addition, for each value of $r$ and $z$, we have an $S^2$ and an $S^5$ with radii that depend on $(r,z)$. The $S^5$ shrinks to zero size on the $r=0$ axis, while the $S^2$ shrinks to zero size at the locations of the conducting plates, so we have various non-contractible $S^3$s and $S^6$s corresponding to paths that terminate on different plates or on different segments of the vertical axis respectively. This is illustrated in figure \ref{fig1}. As shown in \cite{lm}, through an $S^6$ corresponding to a path surrounding plates with a total  charge of $Q$, we have $N_2 = Q$ units of flux from the dual of the Ramond-Ramond four-form, suggesting the presence of $N_2$ D2-branes. Similarly, through an $S^3$ corresponding to a path between plates separated by a distance $d$, we have $N_5 = d $ units of H-flux, suggesting that this part of the geometry between the plates is describing the degrees of freedom of $N_5$ NS5-branes.

Since the matrix model is a massive deformation of the maximally supersymmetric quantum mechanics describing low-energy D0-branes in flat-space, we should expect that the dual supergravity solutions correspond to infrared modifications of the near-horizon D0-brane geometry. Indeed, the solutions are asymptotically the same as the near-horizon D0-brane solution, with the strong-coupling region in the infrared replaced by smooth topological features that depend on the choice of vacuum.

\section{Coarse-Graining the Lin-Maldacena Geometries}

For large $N$, the plane-wave matrix model has of order $\exp(\sqrt{6N}/\pi)$ independent vacua labelled by reducible dimension $N$ representations of $SU(2)$. In this section, we will argue that as for standard thermodynamic systems (e.g. particles in a box), if we use coarse-grained, macroscopic variables to describe the states, then despite the large number of possible microscopic states, the description of a randomly chosen microstate will, with very high probability, be extremely close to the average or ``thermal equilibrium'' state. We will see explicitly what the coarse-grained description of this average state is in our case, and see that there is a natural way to associate a geometry to this (and more general) coarse-grained configurations. We will interpret the resulting geometry as the zero-temperature limit of the thermal state, since this state has a density matrix with equal contributions from each basis vacuum state. Much of the discussion in this section follows ideas in \cite{babel} for the LLM geometries.

\subsection{Macroscopic variables}

We begin by understanding the macroscopic variables appropriate in our case. As we will see, typical gauge theory states for large $N$ will correspond to electrostatics configurations with large numbers of charged disks at unit separation. The microstate configurations are specified by giving the (integer) charge at each discrete location on the vertical axis.  Since the extent of the disk configurations on this axis will be much larger than the disk separations (typically by a factor of $\sqrt{N}$ as we will see), it is sensible to characterize configurations by a macroscopic charge density $Q(z)$. This, we can define by averaging the microscopic charge over a distance much larger than the disk separations, but much smaller than the vertical extent of the disk configuration.  Thus, in the coarse-grained description of states, $Q(z)$ should be a smooth function.

We still need to understand how the charge $Q(z)$ should be arranged in the directions perpendicular to $z$ (recall that for the microstates it spreads out dynamically on the charged conducting disks), but first it will be helpful to see what $Q(z)$ looks like for typical states.

\subsection{Typical states}

In the microscopic description, the charges $Q_n$ at position $z=n$ label how many times the irreducible representation of dimension $n$ appears, and are subject to the constraint
\be
\label{constraint}
\sum_{n=1}^\infty n Q_n = N \; .
\ee
We would now like to ask what a typical randomly chosen representation looks like. To do this, we first note that the independent vacuum states of the matrix model are in one-to-one correspondence with the quantum states of a free massless boson on an interval (a.k.a.\ a quantum guitar string) with energy $E-E_0 = \hbar \omega N$, where $\omega$ is the frequency of the lowest mode. In this analogy, $Q_n$ give the number of particles of frequency $n \omega$. For large $N$, where the energy and number of particles are large, we know that a thermodynamic description is appropriate, and that any macroscopic quantities evaluated for a randomly chosen microstate are extremely likely to be extremely close to the average values.

For our discussion, we will be interested in the average coarse-grained charge distribution defined above, so we start by computing the expected value of $Q_n$ for each $n$. This is equivalent to calculating the expected particle numbers for our gas of free bosons in the microcanonical ensemble at energy $E=N$ (setting $\hbar=\omega=1$). For large $N$, this should agree up to tiny corrections with the result as computed in the canonical ensemble, so long as we choose the temperature such that the expected value of the energy is $N$. The calculation is much simpler in the canonical ensemble, since now we can sum over all states without a constraint.

To study the canonical ensemble, we write a partition function \cite{Baketal}
\bea
Z &=& \sum_{Q_n} e^{-\beta \sum n Q_n} \cr
&=& \prod_n \sum_{Q_n} e^{-\beta n Q_n} \cr
&=& \prod_n {1 \over 1 - e^{- \beta n}} \; .
\label{partfn}
\eea
From this, the expectation value of $Q_n$ is found (for example by changing the $\beta$ in front of $Q_n$ to $\alpha$, differentiating $\ln(Z)$ with respect to $- \alpha n$, and setting $\alpha=\beta$) to be
\be
\label{expQ}
\langle Q_n \rangle = {1 \over e^{\beta n} - 1 } \; .
\ee
The expected value of energy is
\beas
\langle N \rangle = - \partial_\beta \ln(Z) &=& \sum_n {n \over e^{\beta n} - 1} \cr
&\approx& {\pi^2 \over 6 \beta^2} \; ,
\eeas
where the last line assumes that the sum can be approximated by an integral (valid for large $N$). Solving for $\beta$ in terms of $N$ and plugging in to (\ref{expQ}), we find
\be
\label{expQ2}
\langle Q_n \rangle = {1 \over e^{{\pi n \over \sqrt{6 N}} } - 1 } \; .
\ee

Thus, the coarse-grained approximation to a typical microstate will have a linear charge density very close to
\be
\label{expQ2a}
\langle Q(z) \rangle = {1 \over e^{{\pi z \over \sqrt{6 N}} } - 1 } \; .
\ee
Or, defining $x = z/\sqrt{N}$ and $\sqrt{N} q(x)$ to be the charge density in terms of $x$, we have
\be
\label{expQ3}
\langle q(x) \rangle = {1 \over e^{{\pi x \over \sqrt{6}} } - 1 } \; .
\ee

\subsection{Supergravity solution for the average state}

We would now like to understand the supergravity solution corresponding to the average coarse-grained configuration we have found. To do this, we first need to understand precisely how the charge $Q(z)$ should be distributed in the horizontal directions. For the microstates, the actual distribution of charge is determined dynamically, since the charges are free to move on conducting disks whose radii are determined by the constraint that the charge density at the edge vanishes. However, we will now see that the typical configurations for large $N$ with fixed $\lambda$ have disks whose radii are much smaller than the separation between the disks. Thus, in the coarse-grained picture for typical states, we can take the charge distribution to sit on the vertical axis.

To understand how large the disks should be, we note that for the microstates, having conducting disks with the correct radii is necessary in order to avoid singularities in the supergravity solution. If we simply place all the charge on the axis, singularities should appear (wherever $\partial_r V = 0$). These cannot be at radii much larger than the original radii of the disks, since at these large radii, the electrostatics potential should be modified only slightly when we move all the charge to the axis. Thus, the distance scale defined by the sizes of the disks should be the same as the typical coordinate distance from the axis where singularities appear in the modified configuration. We will now use this to estimate the radii of the disks for the typical configurations.

For a charge distribution $Q(z)$ on the vertical axis, the corresponding potential will be given by \cite{Baketal}
\be
\label{axis}
V(r,z) = V_0 (r^2 z - {2 \over 3} z^3) + \int_0^\infty dz' Q(z') \left\{ {1 \over \sqrt{r^2 + (z-z')^2}} - {1 \over \sqrt{r^2 + (z + z')^2}} \right\} \; ,
\ee
where the second term arises from the image charges below the infinite conducting plate. It is straightforward to check that such a potential for smooth $Q(z)$ always gives rise to a singular supergravity solution \cite{Baketal}. The singularity appears at the locus of points where the radial component of the electric field vanishes \cite{lm}. To estimate this radius, we note that for slowly varying $Q(z)$, the radial electric field near the axis is given by
\[
E_r (r) = -2 r z V_0 + 2{Q(z) \over r} \; ,
\]
so the singularity is located at\footnote{This should be a good approximation so long as $r$ is small compared with $Q/Q'$.}
\be
\label{singrad}
r = \sqrt{Q(z) \over z V_0} \; .
\ee

From (\ref{expQ2a}), we see that for $z$ of order $\sqrt{N}$, the typical value of the charge on each disk is of order one, while for $z$ of order one, the typical charge is of order $\sqrt{N}$. Recalling that $V_0 \sim 1/g^2$, we estimate that the typical radii of the disks will be
\beas
r \sim \sqrt{\lambda \over N^{3 \over 2}} \qquad z = {\cal O}(\sqrt{N}) \; , \cr
r \sim \sqrt{\lambda \over N^{1 \over 2}} \qquad z = {\cal O}(1) \; .
\eeas
In either case, for large $N$ and fixed $\lambda$ the typical radii go to zero. Thus, in the coarse-grained description of typical states in the 't Hooft limit, we can take all the charge to be located on the $z$-axis. This leads us to the following conclusion:
{\it the geometry dual to the $T =0$ thermal state of the plane-wave matrix model at large $N$ is given by the Lin-Maldacena solution (\ref{IIA ansatz last}), with potential (\ref{axis}) determined in terms of the charge distribution (\ref{expQ2a}).} It may be that for some coordinate choice, the solution takes a simpler, more explicit form, but we have not investigated this.

\begin{figure}
\centering
\includegraphics{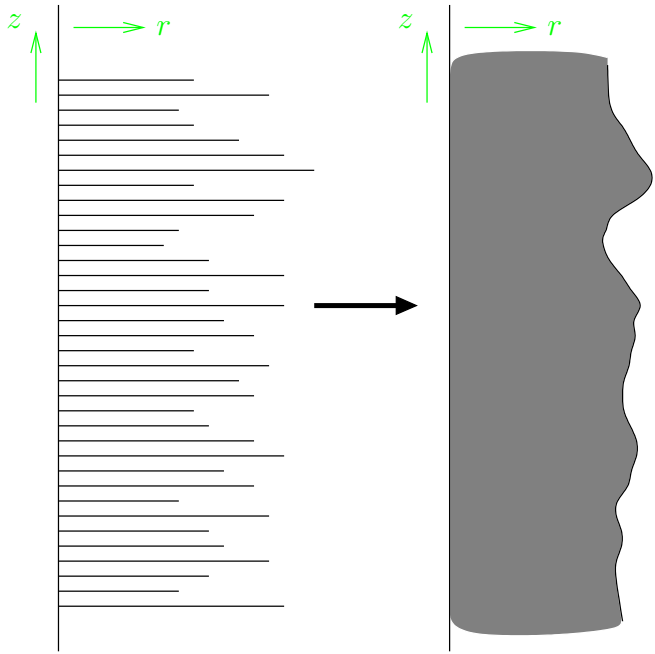}
 \caption{Coarse-graining for large disks. The shaded region represents a solid conductor that conducts only in the horizontal directions.}
\label{fig2}
\end{figure}

\subsection{Coarse-grained solutions for large disks}

For large $N$ and fixed $\lambda$, we have seen that the typical states have electrostatics configurations for which the disks are small relative to their separations, so that the charge can simply be taken to lie on the vertical axis in the coarse-grained description. However, it is also useful to have a coarse-grained description of states in cases where the radii of the disks is larger than their separations. This is relevant, for example, if we allow $\lambda$ to scale as a power of $N$, or for fixed $\lambda$ in restricted ensembles for which we restrict the number of fivebranes (as in section 5).

In such cases, the coarse-grained picture will have the closely spaced disks replaced by a uniform material that conducts only in the directions perpendicular to the $z$-axis. This material will have some smooth profile described by a radius function $R(z)$ and carry charges such that total charge on the conductor between heights $z$ and $z+dz$ is $Q(z)$. Just as the radii of the disks in the original setup are determined by the charges, we should expect that $R(z)$ in the coarse-grained situation will be determined by $Q(z)$. Specifically, it turns out that the shape $R(z)$ of the conductor must be chosen such that the surface charge density vanishes. This $R(z)$ gives the coordinate location of the singularity in the supergravity solution corresponding to a given coarse-grained $Q(z)$. The details of this coarse graining procedure and the mathematical procedure that determines $R(z)$ in terms of $Q(z)$ are described in appendix B.

\section{An entropy functional}

In thermodynamic systems, we can often associate an entropy with coarse-grained configurations that are more general than the state of thermal equilibrium for the whole system. In this section, we give a functional that associates an entropy to a general coarse-grained Lin-Maldacena geometry and discuss its properties. A similar entropy functional has been derived recently for the LLM geometries in \cite{emp,vijaynew}.

\subsection{A familiar example}

As a familiar example, consider an ideal monatomic gas in a box. For a given energy $E$, we can find the entropy of the whole system, but we could also talk about the entropy of a state where all the particles are in one half of the box (but are otherwise in a typical configuration). More generally, we can associate an entropy to an arbitrary configuration for which we specify the particle density and energy density (the macroscopic variables) as a function of position, as long as these vary only over macroscopic scales.

For illustrative purposes, we will work out this example, starting with quantities as calculated in the canonical ensemble. Up to an additive constant, the entropy for $N$ particles in thermal equilibrium at temperature $T$ in volume $V$ is given by
\[
S = Nk(\ln(V/N) +{3 \over 2} \ln(T)) \; .
\]
On the other hand, the average energy is
\[
E = 3/2 N k T \; .
\]
Defining the particle density $\rho$ and the energy density $\rho_E$, we can then write an expression for an entropy density in terms of $\rho$ and $\rho_E$ as
\[
s = S/V = -\rho \ln(\rho) + {3 \over 2} \rho \ln({2 \over 3} \rho_E/k) \; .
\]
Finally, the entropy associated with some general coarse-grained state is
\[
S[\rho, \rho_E] = \int dV \left\{-\rho \ln(\rho) + {3 \over 2} \rho \ln({2 \over 3} \rho_E/k) \right\} \; ,
\]
subject to the constraints that
\[
\int dV \rho = N \; ,
\]
and
\[
\int dV \rho_E = E \; .
\]
We can check that the entropy functional is maximized subject to the constraints for constant $\rho$ and $\rho_E$.

Thus, to define the entropy functional, we split the system up into macroscopic parts (the volume elements), determine the entropy for each of these parts as a function of the coarse-grained variables of the part, and then write the entropy of the whole system as a sum of the individual entropies, with the constraint that the coarse-grained variables are consistent with any specified global quantities (such as energy).

\subsection{Entropy for coarse-grained matrix model vacua}

Now we move on to the plane-wave matrix model vacua. In this case, the variable that we use to describe our coarse-grained configurations is the charge density $q(x)$ (recall that we defined $x = z/\sqrt{N}$. Let us now consider the interval $[x,x+dx)$ as a subsystem of our analog thermodynamic system. The charge in this interval, $q(x)dx$ is given as a sum of independent microscopic variables
\[
q(x)dx = Q_{n} + \dots + Q_{n+l} \; ,
\]
which are also independent of the variables that determine $Q$ outside the interval. Here $n = x \sqrt{N}$ and $l = dx \sqrt{N}$. We assume that the coarse graining is over macroscopic distances, in other words that the number $l$ of individual degrees of freedom contributing to $Q(x)dx$ is large. Thus, we should have $1 \ll l \ll n$. Now, for the subsystem, we have the partition function
\[
Z = \prod_{k=n}^{n+l} {1 \over 1-e^{-\beta k}} \; .
\]
This gives free energy
\[
F \approx l T \ln(1-e^{-n \beta}) \; ,
\]
and energy
\beas
\bar{E} &=& \langle n Q_n + \dots + (n+l) Q_{n+l} \rangle \cr
&\approx& {n l \over e^{n \beta} -1} \; .
\eeas
The entropy is then
\[
S = (E-F)/T = l \left[ {\n \beta \over e^{n \beta} - 1} - \ln(1 - e^{- n \beta}) \right] \; .
\]
Note that this is proportional to the size of the interval, so it makes sense to define an entropy density $s(z) = S/l$ or equivalently $s(x) = \sqrt{N} S /l$. We would like to express this in terms of the average charge density $Q(x)$ in the interval, given by
\beas
Q &=& \langle Q_n + \dots + Q_{n+l} \rangle/dx \cr
&\approx& \sqrt{N}\bar{E}/(n l) \cr
&=& \sqrt{N} {1 \over e^{n \beta} - 1} \; .
\eeas
Solving for $\beta$ in terms of $Q$, and substituting into the formula for $s$, we find
\[
s(x) = \sqrt{N}((q+1) \ln(q+1) - q \ln(q)) \; ,
\]
where we have defined $q = Q/\sqrt{N}$.

Thus, we can associate to a coarse-grained configuration described by a charge density $q(x)$ an entropy
\be
\label{entrop}
S[q(x)] = \sqrt{N} \int dx [(q+1) \ln(q+1) - q \ln(q)] \; .
\ee
Allowed vacua of the matrix model are subject to the constraint
\be
\label{constr}
\int dx x q(x) = 1 \; .
\ee

We can now check that maximizing (\ref{entrop}) subject to the constraint (\ref{constr}) gives the correct result for the charge density. Introducing a Lagrange multiplier for the constraint and varying with respect to $q$, we find
\[
\ln(q+1) - \ln(q) + \Lambda x = 0 \; .
\]
This gives
\[
q(x) = {1 \over e^{\Lambda x} -1} \; ,
\]
and enforcing the constraint yields
\[
\Lambda = \pi/\sqrt{6} \; .
\]
Thus, we reproduce (\ref{expQ3}).

For more general coarse-grained configurations, it is clear from (\ref{entrop}) that the entropy will be nonzero if there is any interval $(x_1,x_2)$ for which $q(x)$ is continuous and nonzero. Thus, the only way to have a vanishing entropy functional with a nonzero net charge is to have the charge located at discrete points on the axis such that $q(x)$ is a sum of delta functions, as we have in the microstate configurations.\footnote{Technically, such a $q(x)$ can only appear as a coarse-grained configuration in the limit where we take the coarse-graining scale to zero. Thus, for any non-zero  coarse-graining scale, the entropy will be non-zero for all configurations.} In this case, the entropy vanishes since for large $q$, we have
\[
(q+1) \ln(q+1) - q \ln(q) \sim \ln(q) \qquad \qquad ({\rm large} \; q)
\]
and
\[
\int \ln(\delta(x-a))dx = 0 \; .
\]
Recalling that the D2- and NS5-brane fluxes are quantized properly in the supergravity solutions if and only if the charges are quantized and located at integer values of $z$, we conclude that the entropy function is zero if and only if $q(x)$ corresponds to a microstate geometry. Consequently, all coarse-grained configurations with non-zero entropy correspond to singular supergravity solutions.

Our formula (\ref{entrop}) gives the entropy as a simple expression in terms of $q(x)$, which in turn directly determines the geometry. In this sense, it is a geometrical formula for the entropy. We might also ask whether there is any direct relation to a horizon area (or Wald's generalization \cite{wald}) in this case. However, as is typical in examples with a large amount of supersymmetry, the singular coarse-grained geometries that we obtain have no horizons.\footnote{ It was shown in \cite{lm} that the  metric components in a general LM geometry \ref{IIA ansatz last} will be continuous and nonzero (except for points on the conducting disks ) for all potential V satisfying the three dimensional Laplace equation.  From this it is straightforward to see that the region outside of the coarse-grained conducting disks is causally connected.}  On the other hand, both the curvature and the dilaton diverge at the singularities, so the supergravity solution should receive both $\alpha'$ and string loop corrections. It is possible that the fully corrected solutions have horizons.

Following \cite{sen95}, we might hope that an appropriate definition of a stretched horizon around the singularity\footnote{Possible definitions considered in the literature include the locus of points where the curvature becomes strong, where the dilaton becomes strong, where the local temperature equals the Hagedorn temperature, or where microstates begin to differ significantly from each other.} would have area that reproduces the entropy (perhaps up to numerical factors). In fact, our setup should provide a very stringent test of any proposed definition of a stretched horizon, if we demand that it correctly reproduces the functional dependence of the entropy on $q(x)$. Unfortunately, as we show in appendix C, the necessary location of a stretched horizon whose area would reproduce our entropy is parametrically closer to the singularity than either the radius where the curvature becomes large or the radius where the dilaton becomes large. At this scale, it is probably naive to expect that a simple area would reproduce the entropy.

\section{Other ensembles}

The $T=0$ thermal solution we have found is analogous to the `hyperstar' geometry of \cite{babel}, dual to the coarse-grained typical state of ${\cal N}=4$ SUSY Yang-Mills theory on $S^3$ with a $U(1) \in SO(6)$ R-charge equal to energy. For that theory, there is a related geometry known as the `superstar' that has been understood as the geometry dual to the equilibrium state in a more restricted ensemble for which the number of D-branes in the spacetime is fixed. There are similar restricted ensembles that are natural to consider in our case.

To understand these, we recall that the microstate geometries contain various non-contractible $S^3$ cycles carrying NS5-brane flux and non-contractible $S^6$ cycles carrying D2-brane flux. For a given microstate, there will be some 3-cycle in the geometry carrying a maximal number of units $N_5$ of NS5-brane flux and some 6-cycle carrying a maximal number of units $N_2$ of D2-brane flux, as shown in figure 3. We loosely refer to $N_5$ and $N_2$ as the number of NS5-branes and D2-branes in the geometry. Just as we understood the typical states in general, we can also ask about the form of the typical states in ensembles where either $N_2$ or $N_5$ or both are fixed.

\begin{figure}
\centering
\includegraphics{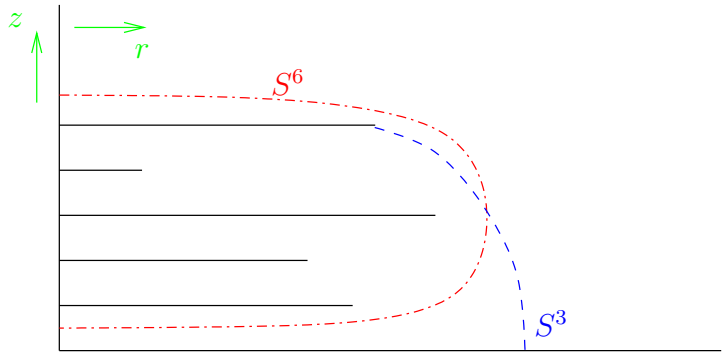}
 \caption{Example electrostatics configuration showing the non-contractible cycles $S^3$ and $S^6$ carrying the largest amount of NS5-brane and D2-brane flux respectively. }
\label{fig3}
\end{figure}

To do this, we note that the total number of units of NS5-brane flux is given by the largest $j$ for which $Q_j \ne 0$, while the number of units of D2-brane flux is given by the total charge $\sum_j Q_j$. If we consider a Young diagram with $Q_j$ rows of length $j$, then $N_2$ and $N_5$ are the total number of rows and columns in the Young diagram respectively. The problem of studying typical Young diagrams with a fixed number of rows (or equivalently a fixed number of columns) is precisely the one studied in \cite{babel} to understand typical states in the hyperstar ensemble of LLM geometries, while the problem of studying typical Young diagrams with a fixed number of rows and columns is precisely the one studied in \cite{babel} to determine the typical configurations in the (generalized) superstar ensemble. Thus, we can directly carry over those results to find the $q(x)$.

\subsection{Fixed $N_5$}

For fixed $N_5$, we simply restrict the partition function (\ref{partfn}) to $n \le N_5$. The expected value of $Q_n$ is given by the same formula,
\be
\label{expQb}
\langle Q_n \rangle = {1 \over e^{\beta n} - 1 } \; ,
\ee
but now the expected value of $N$ is
\beas
\langle N \rangle &=& \sum_{n=1}^{N_5} {n \over e^{\beta n} - 1} \cr
&\approx& N_5^2 f(\beta N_5) \; , \qquad \qquad \qquad f(x) \equiv {1 \over x^2} {\rm Li}_2 (1-e^{-x}) \; .
\eeas
Thus, we obtain a charge density
\[
Q(z) = {1 \over e^{{z \over N_5} f^{-1}(N/N_5^2)} - 1 } \; , \qquad \qquad z \le N_5 \; .
\]
Note that in the unrestricted ensemble, the typical extent of the charge distribution was of order $\sqrt{N}$, so we only have a significant difference from the unrestricted ensemble when $N_5$ is of order $\sqrt{N}$ or smaller. One interesting case is that where we fix $N_5$ to some large but finite value in the large $N$ limit. In this case, we find
\[
\beta = {N_5 \over N} \; ,
\]
and
\[
Q(z) \approx {N \over N_5 z} \; .
\]
In this case, our estimate (\ref{singrad}) for the size of the disks gives $r \sim \sqrt{\lambda/N_5^3}$, so the disks are large compared to their separations for $\lambda \gg N_5^3$. In this case, we need to use the methods of appendix B to determine the appropriate coarse-grained geometry.

\subsection{Fixed $N_2$ or fixed $N_2$ and $N_5$}

For fixed $N_2$ (with either fixed or unrestricted $N_5$), it is simplest to work in a grand canonical ensemble where we introduce a chemical potential for $N_2$ and tune it to get the correct value. We will therefore consider the partition function
\begin{equation} \label{Zgc}
Z(\beta,\mu) = \sum_{Q_j} e^{-\sum (\beta j + \mu) Q_j} .
\end{equation}
From this, we obtain a charge distribution
\be
\label{expQb2}
\langle Q(z) \rangle = {1 \over e^{\beta z + \mu} - 1 } \; ,
\ee
where $\beta$ and $\mu$ are fixed by demanding
\begin{equation}
N = \left< \sum_j j Q_j \right> = \sum_{j=1}^{N_5} {j \over e^{\beta j+\mu}-1}
\; ,
\end{equation}
as before, and
\begin{equation}
\left< \sum_j Q_j \right> = \sum_{j=1}^{N_5} {1 \over e^{\beta j + \mu} -1} \; .
\end{equation}
In general, $\beta$ and $\mu$ are complicated functions of $N_2$ and $N_5$, but as pointed out in \cite{babel}, there is a simple special case where we take $\beta \to 0$ with fixed $\mu$. This gives the solution in the case where we restrict
\[
N_2 N_5 = 2 N \; .
\]
In this case, the charge density is constant
\[
Q(z) = {N_2 \over N_5} \; , \qquad \qquad 0 \le z \le N_5,
\]
and the supergravity solution may be written very explicitly in terms of ordinary functions. This case corresponds to a triangular Young diagram, which in the LLM case gives rise to the original superstar geometry.

We also get a simple expression for the charge distribution in the case where $N_2$ is large but fixed in the large $N$ limit with $N_5$ unrestricted. In this case, a straightforward calculation gives
\[
Q(z) = {N_2^2 \over N_5} e^{-{N_2 \over N_5}z} \; .
\]

\section{Higher dimensional $SU(2|4)$ symmetric theories}

So far, we have discussed the Plane-Wave Matrix Model. However, Lin and
Maldacena \cite{lm} also identified supergravity duals to the vacua of other,
higher dimensional, field theories with $SU(2|4)$ supersymmetry. These are the aforementioned maximally
supersymmetric Yang-Mills theory on $R\times S^2$ \cite{msv}, $\Ncal=4$ SYM
theory on $S^3/Z^k$, and type IIA Little String Theory on $S^5$
\cite{msv,lm,lmsvv}. Aspects of the relations among these theories and the
Plane-Wave Matrix Model have been discussed in \cite{itt,istt}.

In this section we will analyze these theories in the same way as we have for the Plane-Wave Matrix Model. For the higher-dimensional theories, the construction of dual supergravity solutions differs only in the boundary conditions for the electrostatics problem. The individual microstates are still distinguished by the locations and charges of finite-sized conducting disks, so the coarse-graining procedure and the entropy functional are exactly the same as in the Plane-Wave Matrix Model.

\subsection{Maximally supersymmetric Yang-Mills theory on $S^2 \times R$}

\subsubsection{Field theory}

We will first consider maximally supersymmetric field theory on
$S^2 \times R$. This theory can be derived as a limit of the Plane-Wave Matrix
Model \cite{msv}, or of $\Ncal =4$ SYM on $S^3/Z_k$ in the limit
$k\to\infty$ \cite{lm}.

The field content of this theory is the same as the usual low-energy D2-brane gauge theory, with an $SU(N)$ gauge field together with fermions and seven scalar fields.
Six of the scalar fields are associated with the $SO(6)$ R-symmetry of the
theory. The remaining one comes from the dimensional reduction when
the $k\to\infty$ limit is taken in $\Ncal =4$ SYM on $S^3/Z_k$. We will refer
to this scalar as $\Phi$. The vacua of this field theory are given by
$\Phi = -\diag(n_1,n_2,\ldots,n_N)$, and $F=dA=\Phi \sin\theta d\theta d\phi$,
where the $n_i$ are integers, and $\theta$ and $\phi$ are the usual coordinates on $S^2$.

The different vacua of the theory are labelled by the multiplicities of the
integers in the vacuum configurations of $\Phi$ and $F$.

\subsubsection{Supergravity}

The supergravity dual to this theory shares many similarities with the dual
to Plane-Wave Matrix Model. As in the Plane-Wave Matrix Model case, the disks
are parallel, circular, and centred at $r=0$, $z=d_i$.  In this case, however,
the auxiliary electrostatics problem has no infinite disks, and the background
potential is given by
\be
\label{vinf2}
V_\infty = W_0(r^2-2z^2) \; .
\ee
As before, non-singular solutions will have disks with radii $R_i$ chosen so that the
charge density vanishes at the edge of each disk.

Corresponding to a vacuum with $N_i$ copies of the integer $n_i$ will be an
electrostatics configuration with disks at positions $d_i=\pi n_i/2$ carrying
charge $Q_i=\pi^2 N_i/8$. The gauge theory parameters are related to the
electrostatics ones as $g^2_\text{YM} \propto 1/W_0$, and $N=\sum N_i$.

In similar fashion to the Plane-Wave Matrix Model case, we can find the
potential for the system with coarse-grained charge density $Q$ to be
\begin{equation}
\label{axis2}
V(r,z) = W_0 (r^2  - 2z^2) + \int_{-\infty}^\infty dz'   {Q(z') \over \sqrt{r^2 + (z-z')^2}}  .
\end{equation}

\subsubsection{Typical states}

As we have described above, the vacua of this theory are labelled by a set of
integers and their multiplicities. Since the integers specifying the vacuum can be arbitrarily large (the only restriction is that the sum of multiplicities is $N$), we have an infinite number of vacua in this case. In the electrostatics picture, this corresponds to the fact that the plates are allowed to sit anywhere on the $z$-axis, with the only restriction that the total charge is $N$. As a result, quantities such as the charge at any location will average to zero, and we cannot see any natural way to define a typical configuration in this case for the unrestricted ensemble.

On the other hand, we do get a well defined thermal configuration in an ensemble where we fix the number of NS5-branes, as in section 5. This corresponds to fixing the separation between the highest and lowest disk. For the $SU(N)$ theory, we should demand also that the sum of integers times their multiplicities is zero, so we end up with a finite set of vacuum states. For coarse-grained typical states, the total charge $N$ will be evenly distributed between the $N_5$ plates, so the coarse-grained charge density will be
\[
Q(z) = {N \over N_5} \; , \qquad \qquad -{N_5 \over 2} \le z \le {N_5 \over 2}
    \; .
\]

Another way to obtain a non-trivial electrostatics configuration is to recall the definition of this theory as a $k\to\infty$ limit of $\Ncal=4$ SYM on $S^3/Z_k$. If we instead take a limit in which $N\to\infty$ and $k\to\infty$ with
$N/k=\xi$ fixed then the resulting theory will have a $T=0$ thermal state arising from the electrostatics potential
$V(r,z)=W_0(r^2-2z^2)-(\pi \xi)/(2)\ln(r)$. The corresponding geometry will have a
string like singularity with entropy density
\begin{equation}
s = \left(1+\xi\right)\ln\left(1+\xi\right)-\xi\ln\xi .
\end{equation}

\subsection{${\cal N}=4$ Yang-Mills theory on $S^3/Z_k$}

\subsubsection{Field theory}

This theory and its vacua can be obtained from $\Ncal=4$ SYM on $S^3$ in the
following manner, as outlined in \cite{lsv}. We can coordinatize the $S^3$
using the metric
\begin{equation}
ds^2_{S^3} = \frac14[(2d\psi+\cos\theta d\phi)^2+d\theta^2+\sin^2\theta d\phi^2]
\end{equation}
where $\theta$ and $\phi$ are the usual coordinates on $S^2$, and $\psi$ is an
angular variable with period $2\pi$. The orbifold is obtained by identifying
$\psi\sim\psi+2\pi/k$. The vacua of the field theory are given by the space of
flat connections, modulo gauge transformations, on $S^3/Z_k$. The orbifold
allows for vacua of the form $A=-\diag(n_1,n_2,\ldots,n_N) d\psi$,
so that $e^{2\pi n_i/k}$ are $k^\text{th}$ roots of unity. This ensures that $A$ has unit holonomy around the full angular direction $\psi$, which is topologically trivial. To label the vacua
uniquely, we will restrict the integers $n_i$ to be on the interval $[0,k)$.

\subsubsection{Supergravity}

In the supergravity picture, the background potential for ${\cal N}=4$
Yang-Mills theory on $S^3/Z_k$ is the same as in (\ref{vinf2}), but the
electrostatics configuration is required to be periodic in $z$ with period
$\pi k/2$.  Even though the background potential is not periodic in $z$, the
part of the potential that determines the charge densities on the disks is.
So the electrostatics solution will have a periodic part that arises from the
charged disks in addition to the background piece.

The periodic arrays of conducting disks are, in turn, related to the vacua of
the field theory. For a vacuum that has $N_i$ repetitions of the integer $n_i$,
the corresponding electrostatics configuration will have a set of charged
conducting disks at positions
$z=\pi n_i/2, \pi(n_i\pm k)/2, \pi(n_i\pm 2k)/2, \ldots$, each carrying charge
$\pi^2N_i/8$. The gauge theory parameters are given in terms of the
electrostatics parameters by $g^2_\text{YM} k \propto 1/W_0$ and $N=\sum N_i$.

Here the potential for the system with coarse-grained charge density $Q$ is
\begin{equation}
\label{axis3}
V(r,z) = W_0 (r^2  - 2z^2) + \int_{-\infty}^\infty dz' {Q(z')  \over \sqrt{r^2 + (z-z')^2}} ,
\end{equation}
where $Q$ has a of period $\pi k/2$.

\subsubsection{Typical states}

Having described the field theory vacua and the corresponding auxiliary
electrostatics configurations, we would like to consider the typical state.

To find the typical configuration in this case, we can use the partition function \eqref{Zgc} with $\beta=0$. We can fix $\mu$ by imposing
\begin{equation}
N = k\frac{1}{e^{\mu}-1} ,
\end{equation}
which means
\begin{equation}
e^{-\mu} = \frac{N}{N+k} ,
\end{equation}
and the typical vacuum will have $q=N/k$.

Up to an overall constant, the electrostatic potential can be found
outside the charge distribution to be
$V(r,z)=W_0(r^2-2z^2)-(\pi N)/(2k)\ln(r)$.
It is singular, and has an entropy of
\begin{equation}
S = k \left(
    \left(1+\frac{N}{k}\right)\ln\left(1+\frac{N}{k}\right)-
    \frac{N}{k}\ln\left(\frac{N}{k}\right)
    \right) .
\end{equation}

\subsection{Type IIA Little String Theory on $S^5$}

\subsubsection{Field theory}

Type IIA Little String Theory on $S^5$ was defined originally by its supergravity dual, found in \cite{lm} and described below. Using this supergravity dual, it has been argued that this theory 
can be defined by particular double-scaling limits of either the Plane-Wave Matrix Model \cite{lmsvv}, the maximally supersymmetric Yang-Mills theory on $S^2 \times R$ or ${\cal N}=4$ Yang-Mills theory on $S^3/Z_k$ \cite{lsv}.

\subsubsection{Supergravity}

In this case, for the theory associated with $k$ fivebranes we have two infinite
conducting plates separated by a distance $k$. As shown by Lin and Maldacena \cite{lm}, we can have a non-trivial potential 
\be
\label{ns5bc}
V(r,z) = {1 \over g_0} I_0\left({r \over k}\right) \sin\left({z \over k}\right)
\ee
between the plates for which the corresponding geometry has an infinitely long throat carrying NS5-brane flux. The parameter $g_0$ is related to the size of the sphere on which the NS5-branes sit, as measured in units of $\alpha'$ (the dimensionful coupling of the Little String Theory).

We can consider adding additional charged conducting disks to this system while keeping the number of units of NS5-brane flux fixed. In the electrostatics picture, this corresponds to adding some number of finite charged conducting disks in the region between the two infinite disks. The disks can sit at positions $d_i = \pi n_i/2$, where the integers $n_i$ are in the interval $[1,k)$, and carry finite charges $N_i$.

\subsubsection{Typical states}

As for the 2+1 dimensional case, the number of vacua here is infinite if we allow arbitrary configurations finite disks in between the infinite conducting plates. However, it is interesting to consider some restricted ensembles.

First, we add some fixed number $N$ of units of D0-brane flux. This requires that
\[
\sum_i i N_i = N \; .
\] 
In this case, the counting problem is identical to that is section 5.1, so we obtain the same typical charge distribution. Of course, the supergravity solution will be different here, since the background potential is now (\ref{ns5bc}).

Alternatively, we could consider an ensemble of geometries in which the number of units of D2-brane charge is fixed. In that case it is again convenient to
use \eqref{Zgc} with $\beta=0$.  Fixing the asymptotic charge we find that
\begin{equation}
N_2 = (N_5-1)\frac{1}{e^{\mu}-1} ,
\end{equation}
which can be inverted to give
\begin{equation}
e^{-\mu} = \frac{N_2}{N_2+N_5-1} .
\end{equation}
The typical state will have
\begin{equation}
\left< Q_j \right> = \frac{1}{e^{\mu}-1} = \frac{N_2}{N_5-1} ,
\end{equation}
and the entropy of this configuration is, for $N_5\gg1$,
\begin{equation}
S = N_5 \left(
    \left(1+\frac{N_2}{N_5}\right)\ln\left(1+\frac{N_2}{N_5}\right)-
    \frac{N_2}{N_5}\ln\left(\frac{N_2}{N_5}\right)
    \right) .
\end{equation}

\section*{Acknowledgements}

We are grateful to Henry Ling, and especially Vijay Balasubramanian for many useful discussions. This work has been supported in part by the Natural Sciences and Engineering Research Council of Canada, the Killam Trusts, the Alfred P. Sloan Foundation, and the Canada Research Chairs programme.

\appendix

\section{The Lin-Maldacena solutions} \label{sgapp}

The general Lin-Maldacena $SU(2|4)$-symmetric supergravity ansatz (suppressing an overall factor of $\alpha'$ in the metric) is given by \cite{lm}
\begin{eqnarray}
ds_{10}^{2} &=& \left( \ddot V - 2 \dot V \over - V'' \right)^{1/2}
\left\{ - 4 { \ddot V \over \ddot V - 2 \dot V } dt^2 + { - 2 V'' \over \dot V} ( d\rho^2
+ d\eta^2 )  + 4 d\Omega_5^2 + 2{ V'' \dot V \over   \Delta }d\Omega_2^2
\right\}  \; , \cr
e^{4\Phi  } &=&{ 4 ( \ddot V - 2 \dot V)^3 \over - V'' \dot V^2
\Delta^2 } \; ,  \cr
C_{1} &=&- { 2 \dot V' \, \dot V \over \ddot V - 2 \dot V } dt \; , \label{IIA ansatz last} \\
F_{4} &=&d C_3  ,\quad \quad \quad
 \quad \quad C_{3}=- 4  { \dot V^2 V'' \over   \Delta } dt\wedge d^{2}\Omega , \cr
H_{3} &=&d B_2 ~,~~~~~~~~~~~~~B_{2}   = 2 \left( { \dot V \dot V' \over   \Delta} + \eta \right) d^{2}\Omega \; ,
\cr
  \Delta &\equiv& (\ddot V - 2 \dot V) V'' - ( \dot V')^2 \; .
\nonumber
\end{eqnarray}

In these equations, the potential $V$ uses slightly different conventions from the one we discussed. The potential $V_{lm}$ here is related to our potential $V$ by
\[
V_{lm}(r,z) = {\pi \over 4} V( {2 \over \pi}r, {2 \over \pi} z) \; .
\]

\section{Coarse-graining for large disks} \label{largedisk}

For certain parameter values, or in restricted ensembles, the typical states are such that the radii of the disks are large compared to their separations. As we noted above, in this case, the macroscopic description will replace the closely spaced disks with a solid material that conducts only in the horizontal directions.

Such a conductor has the following properties. Since the charges are free to rearrange themselves in the directions perpendicular to $z$, they will do so in such a way that the final potential inside the conductor is a function only of $z$, ensuring that the electric field in the $r$ and $\theta$ directions is zero. There will generally be some charge distribution inside the conductor, given by
\be
\label{chargein}
\rho(z) = -{1 \over 4 \pi} V''(z) \; ,
\ee
so $\rho$ is also a function only of $z$. The remaining charge will build up at the surface of the conductor.
In general, the shape $R(z)$ for the conductor, and the linear charge distribution $Q(z)$ on the conductor, together with some fixed background potential will determine the charge density $\rho(z)$ inside the conductor and the surface charge density $\sigma(z)$, determined from $\rho(z)$ via
\be
\label{netcharge}
Q(z) = \pi R^2(z) \rho(z) + 2 \pi R(z) \sigma(z) \sqrt{1 + (R'(z))^2} \; .
\ee
On the other hand, for some special choice of $R(z)$, the surface charge density will vanish. This is the coarse-grained analogue of the constraint that the charge density should vanish at the tip of the disks.

\subsection{The variational problem}

We will now set up the mathematical problem that determines $R(z)$ and $\rho(z)$ from $Q(z)$. We start by assuming some fixed $R(z)$ and $Q(z)$.

Outside the conductor, the potential will be given by
\[
V_+(r,z) = V_0(r,z) + \tilde{V}(r,z) \; ,
\]
where $\tilde{V}$ is the potential due to the charges in the conductor, which should vanish at large $r$ and $z$. Since $\tilde{V}$ is an axially symmetric solution of Laplace's equation, we can expand it in terms of Bessel functions,
\[
\tilde{V}(z) = \int_0^\infty {du \over u} A(u) e^{-z u} J_0 (r u) \; .
\]
Inside the conductor, the potential will be some function $V_-(z)$. The unknown functions $A(u)$ and $V_-(z)$, together with the charge density $\rho(z)$ inside the conductor and the charge density $\sigma(z)$ on the surface of the conductor will be determined by the two equations (\ref{chargein}) and  (\ref{netcharge}), and the boundary condition
\be
\label{boundary}
\vec{E}_+(R(z),z) - \vec{E}_-(z) = 4 \pi \sigma(z) \hat{n} \; .
\ee

In our case, we wish to fix $R(z)$ by the constraint that the surface charge density vanishes. Then the electric field must be continuous across the boundary of the conductor, and since the electric field is vertical inside, we must have $\partial_r V(R(z),z)=0$. Explicitly, we have
\be
\label{givesR}
\partial_r V_0(R(z),z) - \int_0^\infty du e^{-zu} A(u) J_1(R(z) u) = 0 \; .
\ee
This determines $R(z)$ in terms of $A(u)$. Given this, the potential inside the conductor is determined by the $z$ component of the boundary condition (\ref{boundary}), or simply by continuity of the potential across the boundary, so
\[
V_-(z) = V_0(R(z),z) + \int_0^\infty {du \over u} A(u) e^{-z u} J_0 (R(z) u) \; .
\]
Finally, we can use (\ref{chargein}) and (\ref{netcharge}) to write an equation relating $A(u)$ and $Q(z)$,
\be
\label{givesA}
Q(z) = -{1 \over 4}  R^2(z) (\partial_z^2 V_0(z) + \int_0^\infty du u A(u) e^{-z u} J_0 (R(z) u)) \; .
\ee

To summarize, $A(u)$ is determined by the integral equation (\ref{givesA}) where $R(z)$ is determined in terms of $A$ via (\ref{givesR}).

In practice, it is far simpler to determine $R(z)$ and $Q(z)$ given some $A(u)$, or more generally some solution to the Laplace equation that arises from any set of axially symmetric localized charges. We could also parametrize our solution to the Laplace equation via the multipole data rather than the function $A(u)$. As an example of this approach, we can come up with an explicit coarse-grained supergravity solution starting with the simplest non-trivial solution $\tilde{V}$, namely the potential from a dipole localized at the origin (the infinite conducting plane at $z=0$ forces the potential to be an odd function of $z$.). In this case, we have
\[
\tilde{V}(r,z) = p {z \over (r^2 + z^2)^{3 \over 2}} \; .
\]
The radial electric field for the full potential is then
\[
E_r(r,z) = -\partial_r V_+(r,z) = -2 V_0 r z + 3 p {rz \over (r^2 + z^2)^{5 \over 2}} \; .
\]
Requiring that this is zero gives $r=0$ or $z=0$ or
\[
z^2 + r^2 = x^2 \; ,
\]
where we define
\[
x = \left( {3p \over 2 V_0} \right)^{1 \over 5} \; .
\]
Thus, in this case, the profile of the conductor is spherical. From (\ref{givesA}), we can now determine the corresponding charge density $Q(z)$. We find
\[
Q(z) = {5 \over 2} V_0 z (x^2 - z^2) \; .
\]
As a check, we find that the total dipole moment for this configuration is
\[
\int_0^\infty dz 2z Q(z) = p \; .
\]

So we have at least one example where we know both the geometry and the Young diagram explicitly. Note that for this case, the typical height for the plates and the typical size are the same, of order $x$. In terms of the field theory parameters, we have $V_0 \sim  1/g^2$ and $p=2N$, so $x  \sim  \lambda^{1 \over 5}$. Thus, our coarse-grained description should be valid as long as $\lambda$ is large. The typical charge on one of the plates in the corresponding microstate geometries is $Q  \sim  V_0 x^3  \sim  N/\lambda^{2 \over 5}$. In section 3, we saw that this charge is of order one for typical distributions, so it is only for $\lambda  \sim  N^{5 \over 2}$ that the geometry we have constructed has an entropy of the same order of magnitude as the thermal state. (It is important to note that for a fixed configuration of disks (i.e.\ fixed $p/V_0 \sim \lambda$), the corresponding charge distribution changes as a function of $N$.)

\section{Stretched horizons}

In this appendix we investigate the possibility that the area (or some generalization of area) of a suitably defined stretched horizon might reproduce the entropy formula (\ref{entrop}).\footnote{Recently, these ideas have been explored in the context of coarse-grained LLM microstates \cite{references}, though a prescription for defining a stretched horizon that generally reproduces the entropy of coarse-grained states has not emerged.} We focus on a particularly simple specific example of a coarse-grained geometry, and find that a stretched horizon whose area would reproduce the entropy would necessarily be parametrically closer to the singularity than both the scale $x_s$ where the string coupling becomes of order one, and the scale $x_c$ where the curvature becomes string scale.

The geometry we focus on is the thermal state geometry of the super Yang Mills theory on $S^{3 } /Z_k $. In this case the potential is simply
\be
\label{ potential}
-\frac{N\pi }{2k}\log\rho +V_0 (\rho^2 -2\eta^2)
\ee
where $V_0 \sim \frac{1}{g_{ym}^2 k}$ as identified in \cite{lsv}. The potential is singular at $\rho = 0 $, which violates the regularity condition on the LM geometry. The boundary of the coarse-grained conducting disks is at $\rho =r_0 \sqrt{\tfrac{\pi N}{4kV_0}}$, and the supergravity solution is
\bea \label{geometry}
ds_{10}^{2} &=& \left( N \over 4V_0 k\pi  \right)^{1/2}
\left\{ - 4 { (4V_0 k \rho^2 )\over N\pi } dt^2 + { 8 V_0 \over (2V_0  \rho^2 -N\pi /2k )} ( d\rho^2
+ d\eta^2 )  + 4 d\Omega_5^2 \right.\cr
&+& \left.2{ k (2V_0 \rho^2 -N\pi /2k ) \over   N\pi }d\Omega_2^2 \right\} \cr
e^{4\Phi  } &=& { ( N\pi /k ) \over 16 V_0^3 (2V_0  \rho^2 -N\pi /2k )^{2} } ~,~C_{3}=- 4  { k (2V_0 \rho^2 -N\pi /2 k )^{2} \over  \pi N } dt\wedge d^{2}\Omega ~,~B_{2}   = 2\eta  d^{2}\Omega
\eea
We see explicitly that the geometry is singular at $\rho =r_0 \sim\sqrt{g^2 N}$, which is exactly the edge of the disks, but there is no horizon in this geometry. This solution has been considered in \cite{llm}, where it was pointed out that the singularity is related to the $Z_k $ orbifold singularity in the IIB language.
We will assume the stretched horizon to be a constant $\rho $ surface respecting the translational symmetry along the $\eta $ direction.  Using $\rho =r_0 +x $, we find the string coupling becomes of order one at
\be \label{xs}
x_s = {1 \over 8\sqrt{\pi }V_0^2 }\sim(g_{ym}^2 k)^2 .
\ee
 The Ricci scalar can be calculated noticing the fibred structure of the metric,
\be \label{rscalar}
R^{string}=3\sqrt{\frac{V_0 k }{N\pi }}\frac{8V_0 k\rho^2 -N\pi }{4V_0 k\rho^2 -N\pi } .
\ee
We see that it diverges at exactly the boundary of the coarse-grained conducting disks. The curvature becomes of string scale at
\be  \label{xc}
x_c \sim 1 \; .
\ee
In the above we have assumed $g_{ym}^2 N \gg 1 $ in order for the supergravity approximation to be valid. As a result, we will be interested in the scale where $g_{ym}^2 N \gg 1 \gg g_{ym}^2 k $, and in particular $N/k \gg 1 $.
The area of an 8-surface at constant $t$ and $\rho =R(\eta) $ can be calculated to be (in the Einstein frame)
\be \label{areaone}
A =2^{11/2}\omega_2 \omega_5 \sqrt{1+R'^2(z)}\sqrt{(\ddot{V}-2\dot{V})}\dot{V}^{3/2} ,
\ee
where $\omega_2 , \omega_5 $ are the volume elements of the two-sphere and five-sphere respectively. Specializing to $R (z) =r_0 +x $ and to the metric \eqref{geometry}, we get
\be \label{areatwo}
A=16\omega_2 \omega_5 \sqrt{N\pi } \frac{(4V_0 k \rho^2 -N\pi )^{3/2}}{k^2 } .
\ee
We note that it is a monotonically increasing function with the distance from $r_0 $. Using this and evaluating at $x_c, x_s $ we find
\bea \label{ac}
A_c &\sim& \frac{N^{5/4}}{g_{ym}^{3/2}k^2} \; , \cr
A_s &\sim& (g_{ym}^2 k)^3 A_c \; .
\eea
The Bekenstein-Hawking entropy formula $S = {A \over G_N} $ gives ($G_N =g_s ^ 2 =(g_{ym}^2 k)^2 $)
\bea \label{sc}
S_c &\sim& \frac{1}{g_{ym}^{ 11/2}k^{11/4}}\left(\frac{N}{k}\right)^{5/4} , \cr
S_s & = & (g_{ym}^2 k)^3 S_c .
\eea
As expected $S_c \gg S_s $.
According to the entropy functional \eqref{entrop}, the entropy associated with the geometry \eqref{geometry} is
\be \label{entropy}
S=-k\ln(N/k)+(N+k)\ln(N/k+1) .
\ee
In the large $N/k$ limit it becomes
\be \label{entropyone}
S\sim  k(\ln(N/k)+1) ,
\ee
which is much smaller than both $S_s, S_c $. Here, both $\alpha'$ and string loop corrections are very important. Further, if a horizon (or some stringy analogue) does exist in the fully corrected solution, we may require a highly stringy generalization of area to compare with the entropy. While we have studied only one particular example, we expect that the qualitative features will apply in the general case.

\end{document}